\documentclass[12pt]{iopart}
\usepackage{iopams}
\usepackage{graphicx}

\begin{document}

\title{A scalable optical detection scheme for matter wave interferometry}
\author{Alexander Stibor, Andr\'{e} Stefanov, Fabienne Goldfarb$^1$, Elisabeth
Reiger, and Markus Arndt}
\address{Institut f\"{u}r Experimentalphysik, Universit\"{a}t Wien, Boltzmanngasse 5, A-1090
Wien\\ $^1$Laboratoire Aim\'{e} Cotton, F-91405 Orsay Cedex}
\ead{markus.arndt@univie.ac.at}

\date{\today}

\begin{abstract}
Imaging of surface adsorbed molecules is investigated as a novel
detection method for matter wave interferometry with fluorescent
particles. Mechanically magnified fluorescence imaging turns out
to be an excellent tool for recording quantum interference
patterns. It has a good sensitivity and yields patterns of high
visibility. The spatial resolution of this technique is only
determined by the Talbot gratings and can exceed the optical
resolution limit by an order of magnitude. A unique advantage of
this approach is its scalability: for certain classes of nanosized
objects, the detection sensitivity will even increase
significantly with increasing size of the particle.
\end{abstract}

\maketitle

\section{Introduction}
Recent years have seen a tremendous progress in the development of
various matter wave interference experiments, using
electrons~\cite{Freimund2001a,Ji2003a}, large ultra-cold atomic
ensembles~\cite{Andrews1997a}, cold
clusters~\cite{Schollkopf1994a,Bruehl2004a} or hot
macromolecules~\cite{Arndt1999a,Brezger2002a,Hackermuller2003a}.
Interferometry is also expected to lead to interesting
applications in molecule metrology and molecule lithography. In
particular quantum interference of complex systems is intriguing
as it opens new ways for testing fundamental decoherence
mechanisms~\cite{Hackermuller2004a}.

Further progress along this line of research requires an efficient
source, a versatile interferometer and a scalable detection
scheme. Scalable sources represent still a significant
technological challenge, but an appropriate interferometer scheme
has already been suggested~\cite{Clauser1997a} and successfully
implemented~\cite{Brezger2002a} for large molecules. All coherence
experiments with clusters or molecules up to date finally employed
ion detection. However, most ionization schemes run into
efficiency limits when the mass and complexity of the particles
increases~\cite{Schlag1992a}. Surface adsorption in combination
with fluorescence detection is therefore a promising alternative.
Its high efficiency will reduce the intensity constraints on
future molecular beam sources for interferometry. It appears to be
an important prerequisite not only for experiments exploring
molecular coherence beyond 10,000\,amu but also for decoherence
and dephasing experiments with various dyes below 1,000\,amu.

In the present article we demonstrate the feasibility of optically
detecting matter wave interference fringes of dye molecules. Such
structures usually have periods between $100...1000$\,nm and would
be hardly resolved in direct imaging. But we show that a
mechanical magnification step is a simple and very efficient
technique to circumvent the optical resolution limit for
interferograms. We can thus combine the high sensitivity of the
fluorescence method with the high spatial resolution of our
interferometer setup.

\begin{figure}[tbp]
\includegraphics[width=0.9\columnwidth]{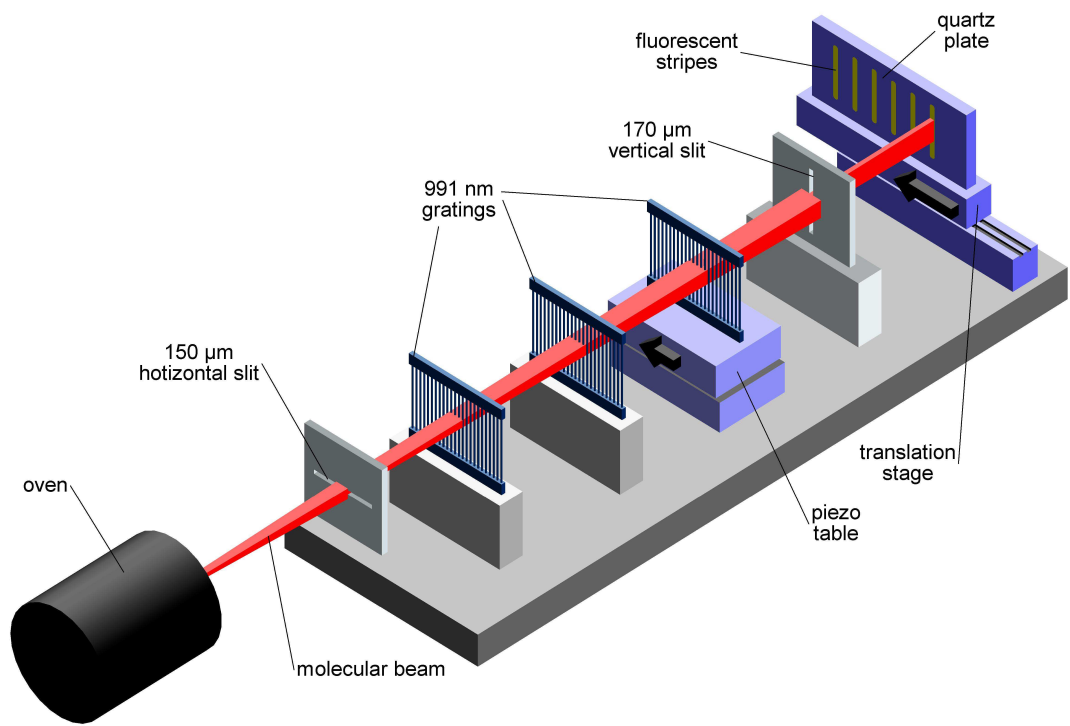}
\caption[setup]{Mechanically magnified imaging of molecular
interference patterns. All molecules passing the third grating are
adsorbed on a plate which scans perpendicular to the molecular
beam. The step-size of the adsorber is 4250 times larger than
that of the third grating which is scanning along the same
direction.} \label{Fig1:Setup}
\end{figure}

\section{Experimental setup}
\subsection{Talbot-Lau interferometry with molecules}
The idea behind the Talbot Lau interferometer has been previously
described for instance in~\cite{Clauser1994a,Dubetsky1997a,
Brezger2002a,Hackermuller2003a} and the modification by our new
detection scheme is shown in Fig.~\ref{Fig1:Setup}. Molecules,
which pass the device, reveal their quantum wave nature by forming
a regular density pattern at the location of the third grating.
The distances between the gratings are equal in the experiment and
corresponds to the Talbot length $L=0.38$\,m of molecules with a
velocity of 250 m/s. It is chosen such that the period of the
molecular fringe system is the same as that of the third grating.
The regular interference pattern can then be visualized by
recording the total transmitted molecular flux as a function of
the position of the transversely scanning third grating.

The quantum wave nature of various large molecules was already
studied in a similar interferometer but using either laser
ionization~\cite{Brezger2002a} or electron impact ionization in
combination with quadrupole mass spectrometry for the detection of
the molecules~\cite{Hackermuller2003a}.  In spite of their
success, both previous methods will probably be limited to masses
below 10,000\,amu. It is therefore important to develop a
scalable detection scheme, such as fluorescence recording, which
does not degrade but rather improve with molecular mass.

Molecule interferometry often operates with low particle numbers.
Direct molecule counting in free flight therefore typically
exhibits too weak signals~\cite{Stefanov2004a}. In contrast to
that, the light-exposure time of surface adsorbed molecules can
exceed that of free-flying particles by orders of magnitude. When
bound to a surface the molecules may also release part of their
internal energy to the substrate. This helps in limiting the
internal molecular temperature and in maximizing the number of
fluorescent cycles. In our experiment the molecular beam was
collected on a quartz plate behind the third grating. Several
studies were done with similar aromatic molecules on silica or
quartz surfaces. Due to their insulating properties the molecular
fluorescence yield exceeds that on simple metals or
semi-conductors~\cite{Tanimura1980a}.

\subsection{Mechanically magnified fluorescence imaging}
For demonstrating the feasibility of this novel detection method
we chose meso-Tetraphenylporphyrin (TPP, Porphyrin Systems
PO890001), a biodye with a mass of 614\,amu. It exhibits
sufficiently strong fluorescence, and a sufficiently high vapor
pressure~\cite{Stefanov2004a} to be evaporated in a thermal source
which was set to a temperature of about 420\,$^{\circ}$C. Moreover
it was known to show quantum wave behavior in our
setup~\cite{Hackermuller2003a}. Quadrupole mass spectroscopy
allowed us to determine a mass purity of approximately 93\%. A
small contribution (7\%) of porphyrin molecules lacked one phenyl
ring. Smaller contaminations may contribute up to 2\% to the mass
in the initial powder but not to the fluorescence on the surface.

The adsorbing quartz surface was mounted on a motorized
translation stage and it was shifted stepwise, parallel to the
third grating as shown in Fig.~\ref{Fig1:Setup}. A fixed slit
between the third grating and the quartz plate  with a width of
170\,$\mu$m limited the exposed area on the surface.

Molecules were deposited and accumulated over a time span of eight
minutes under stationary conditions. Then the third grating, with
a grating period of 991\,nm, (about 400\,nm open slits and a
thickness of 500\,nm) was shifted by 100\,nm, and the adsorber
plate was simultaneously displaced by 425\,$\mu$m to an unexposed
spot.

By repeating this process more than 30\,times the third grating
was moved over three periods and 30\,stripes of fluorescent TPP
molecules were accumulated on the surface. This way the molecular
interference pattern was recorded with a mechanical magnification
factor of 4250. The large (260 $\mu$m) gap between two stripes
prevented any mixing of the molecules which could otherwise be
caused by surface diffusion between the stripes. We have verified
in independent diffusion experiments with TPP on quartz surfaces
that the molecules aggregate and get immobilized on the
300...400\,nm scale at room temperature.

With our new method the resolution is only limited by the
dimensions of the gratings in the interferometer. These may have
openings down to 50\,nm and periods as small as
100\,nm~\cite{Savas1995a} as already used in earlier molecule
interference experiments~\cite{Arndt1999a,Bruhl2002a}.

\subsection{Velocity selection}
High contrast interferences fringes require that the molecular
velocity spread be not too large. Actually for TPP and our present
grating period of 990\,nm a width of $\Delta v/v\simeq 10\%$ is
sufficient. As in earlier experiments~\cite{Brezger2002a}, this is
done by selecting certain free-flight parabola of the molecules in
the Earth's gravitational field using three horizontally oriented
slits. The first slit is provided by the oven aperture of
200\,$\mu$m, another slit of 150\,$\mu $m width is placed 1.2\,m
away from the oven. The third point of the parabola is given by
the vertical position on the detecting surface, which is located
2.9\,m behind the oven. Fast molecules arrive at the top of the
plate. Slow molecules, with a longer falling time, reach the
surface at a lower position. In principle our method therefore
provides the option to select the longitudinal coherence length a
posteriori, after the experiment is already finished. However, in
the present configuration the width of the velocity distribution
is essentially determined by the size of the first two slits,
since we integrate typically only over a much smaller position
interval of 33\,$\mu $m on the surface.

\subsection{Surface preparation and fluorescence readout}
An important requirement for the experiment is to have a perfectly
clean substrate of low self-fluorescence. We used fused silica
(suprasil I) of 500\,$\mu $m thickness. It was cleaned from dust
and organic solvents using the RCA-1 cleaning
procedure~\cite{Kern1970a} followed by methanol sonication and
rinsing with ultrapure water. The clean surface was then bleached
by an expanded 16\,W argon ion laser beam for 30\,minutes with an
intensity of about 3\,W/cm$^{2}$. Owing to this preparation, no
further bleaching could be observed and the background
fluorescence was correspondingly low.

After depositing the molecules, the quartz plate was removed from
the vacuum chamber and put under a fluorescence microscope (Zeiss;
Axioskop 2 mot plus) in air, where a picture of each stripe was
taken with an optical magnification factor of 20 and an
integration time of 20\,s. The irradiation intensity was
1.5\,W/cm$^2$. TPP absorbs well in the blue and emits in the red.
Correspondingly, we used a standard mercury lamp (HBO 100) with an
excitation filter transmitting wavelengths between 405 and
445\,nm, a dichroic beam splitter with a pass band above 460\,nm
and an emission filter which transmitted above 600\,nm.

\begin{figure}[tbp]
\includegraphics [width=1.0\columnwidth ] {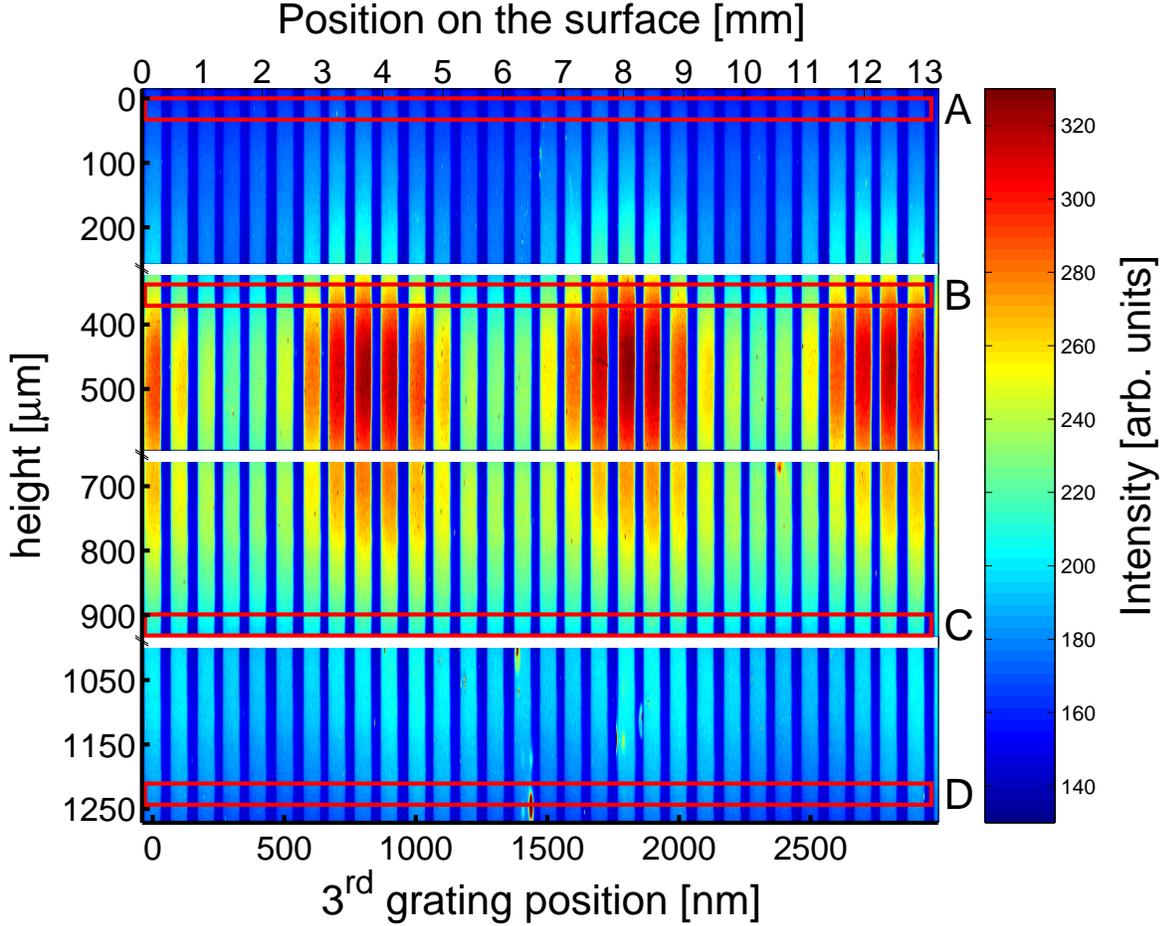}
\caption{False color image of the measured fluorescent molecule
density pattern on the surface. The images were corrected as
described in the text. Each stripe corresponds to a given position
of the third grating, and the vertical distribution encodes the
molecular velocity. Four vertical cuts were made to measure the
interference visibility of molecules with different velocities:
A=33\thinspace $\protect\mu $m (v$\approx $300 m/s), B=371\thinspace $\protect%
\mu $m (v$\approx $270 m/s), C=938\thinspace $\protect\mu $m
(v$\approx $160 m/s) and D=1234\thinspace $\protect\mu $m
(v$\approx $140 m/s).} \label{Fig2:Stripes}
\end{figure}

At the chosen optical magnification and because of the limited
size of the CCD camera (1\,Megapixels) a single microscope image
covers a height of 340\,$\mu$m. As the molecular beam is spread
over about 3000\,$\mu$m due to its velocity distribution, a whole
picture matrix had to be recorded to image all velocity classes.
Four rows of this 9 x 30 matrix selected around the positions
with the highest molecular coverage are shown in
Fig.~\ref{Fig2:Stripes}.

The high quality of the data could be obtained because of the
initial surface preparation and no smoothing was needed. The
single images of one matrix row were arranged a bit closer to each
other than they lie on the surface. Most of the empty gap between
the stripes was removed for presentation purposes and their upper
and lower ends were clipped to avoid regions of optical
aberration.

From available vapor pressure data for porphyrins
~\cite{Stefanov2004a,Perlovich2000a}, we estimate that around
0.1\,monolayers of TPP reach the surface in eight minutes. Based
on the work by~\cite{Tanimura1980a} we assume that the
fluorescence signal grows linearly with the deposition time within
our experimental parameter range. Hence the fluorescence signal is
proportional to the incident intensity, $I_{i}(x,y)$, the
molecular fluorescence efficiency $\eta $, a geometrical
collection factor $K\left( x,y\right) $ and to the molecular
surface number density $N(x,y)$, which we want to determine:
\begin{equation}
I_{f}(x,y)=\left[\eta N(x,y)+B\right] \cdot
K(x,y)I_{i}(x,y)+I_{c}(x,y) \label{fl}
\end{equation}%
where $B$ is the background fluorescence emitted by the
illuminated substrate and $I_{c}(x,y)$ represents the intrinsic
detector noise. The surface sticking coefficient is assumed to be
independent of the molecular surface coverage in our density
regime and is included in N. The intensity of a reference image on
a clean portion of the surface without molecules is
\begin{equation}
I_{r}(x,y)=BK(x,y)I_{i}(x,y)+I_{c}(x,y)  \label{ref}
\end{equation}%
Using Eq.~(\ref{fl}) and Eq.~(\ref{ref}) we can evaluate from the
experimental data the molecular surface density up to a constant
factor
\begin{equation*}
\tilde{N}(x,y)=\frac{\eta }{B}N(x,y)=\frac{I_{f}(x,y)-I_{c}(x,y)}{%
I_{r}(x,y)-I_{c}(x,y)}-1
\end{equation*}%

\begin{figure}[tbp]
\includegraphics [width=0.8\columnwidth] {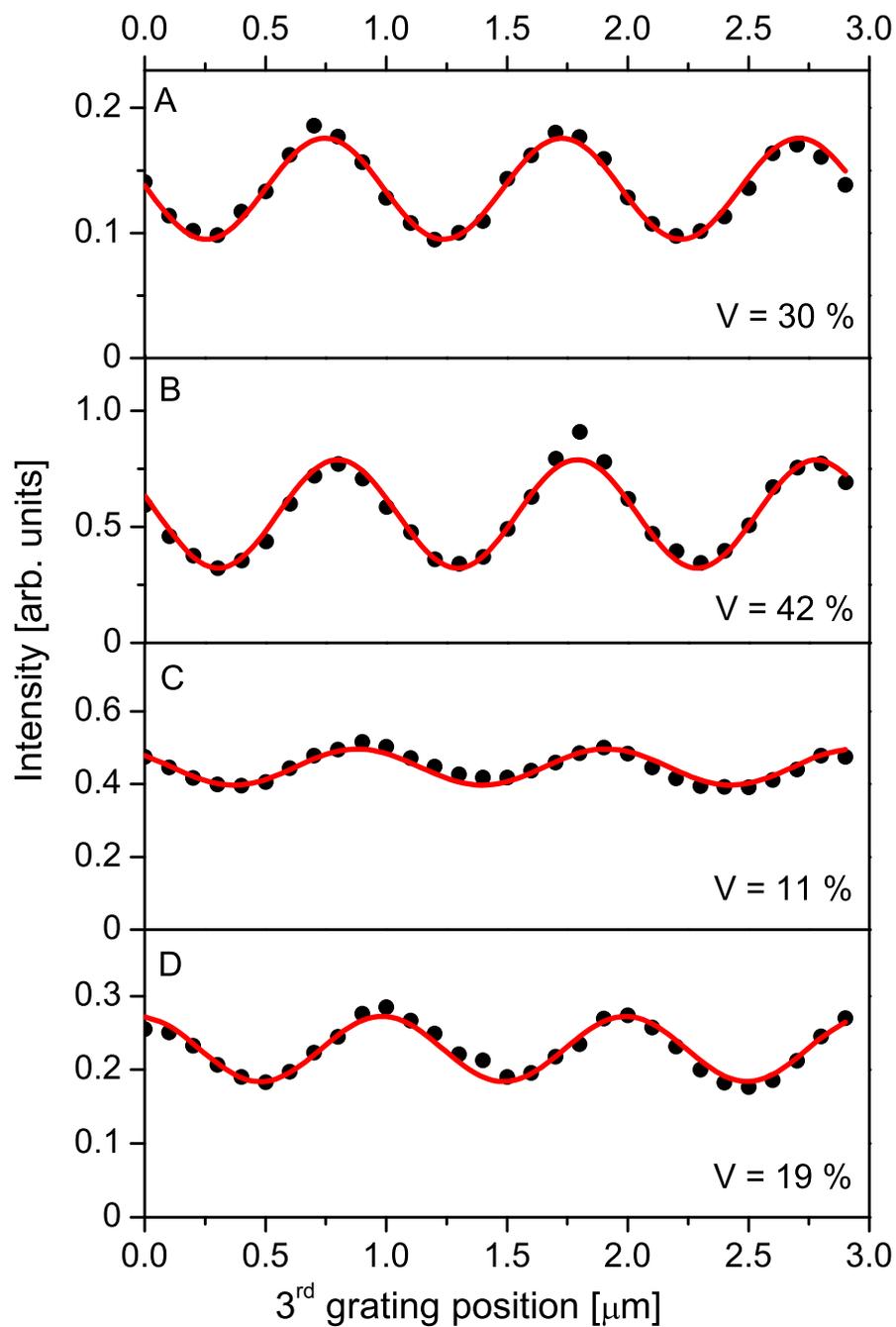}
\caption{Interferences fringes extracted from the fluorescence
images at different heights. The high contrast and its variation
with velocity can only be explained by quantum interference. The
labels A to D refer to Fig.\,2.} \label{Fig3:Fits}
\end{figure}

\begin{figure}[tbp]
\includegraphics [width=0.8\columnwidth] {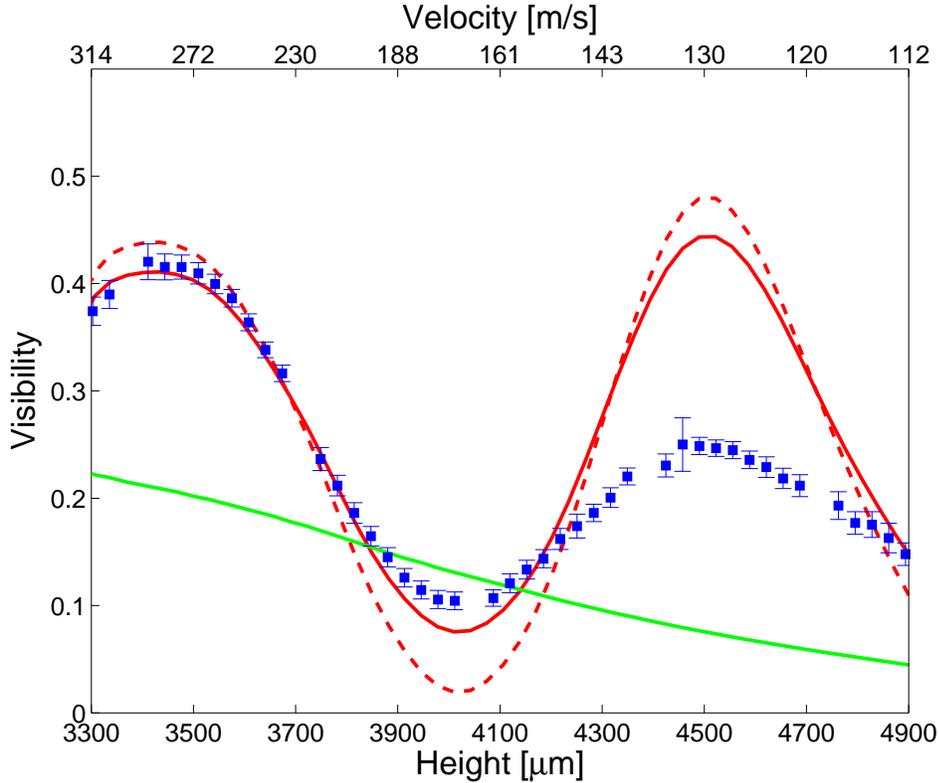}
\caption{Visibility as a function of the deposition height. The
dashed curve is computed by averaging over a velocity distribution
whose mean is shown on the upper axis. The red plain curve is a
corrected theory as described in the text. The green curve shows
the contrast which would be expected by a classical Moir\'{e}
effect. The experiment (full squares) clearly follows the quantum
and not the classical model. Deviations at low velocities are
discussed in the text.} \label{Fig4:Vvsv}
\end{figure}

Fig.~\ref{Fig2:Stripes} shows the corrected intensity distribution
$\tilde{N}(x,y)$. For each vertical stripe the total signal
$\tilde{N}_{tot}(h)$ is computed by integrating $\tilde{N}(x,y)$
over a rectangle centered at position $h$ in the middle of the
stripe. The integration height is $33\,\mu $m and the width is
$100\,\mu $m. The resulting intensity cross sections for four
heights selected in Fig.~\ref{Fig2:Stripes} (A, B, C and D) are
shown in Fig.~\ref{Fig3:Fits}. An evaluation of altogether 43 such
interference curves allows to create a smooth plot of the
interference fringe visibility versus the molecular velocity, as
shown in Fig.~\ref{Fig4:Vvsv}.  Note, that for TPP the velocity
class with the highest contrast (b in Fig.\,2 at h=350 $\mu$m) is
very close but not equal to the most probable velocity (h=500
$\mu$m).

\section{Results}
The experimental fringe visibility (full squares in
Fig.~\ref{Fig4:Vvsv}) clearly varies in a non-monotonic and
quasi-periodic way with the vertical molecular position on the
detector, i.e. with the velocity or the de Broglie wavelength. The
classical model, shown as the falling green line, cannot even
qualitatively reproduce the velocity dependence of the fringe
contrast -- even if we take into account the van der Waals
interaction with the grating walls as done here. The quantum
prediction (dashed curve) also includes the molecule-grating
interaction. It is computed by averaging the theoretical
visibility over the velocity distribution, which is obtained from
the geometry of our setup. The experimental contrast is well
reproduced for fast molecules (about 250\,m/s) and falls below
the quantum model for velocities around 130\,m/s. Slower
molecules are more sensitive to both laboratory noise causing
interferometer vibrations~\cite{Stibor2005a} and to collisional
decoherence~\cite{Hornberger2003a}. The observed contrast
reduction for porphyrins at about 130\,m/s is consistent with
these effects. This is however not a fundamental limit, as in
future experiments the present base pressure of $2\times
10^{-7}$\,mbar in the interferometer chamber can certainly be
improved by about two orders of magnitude. And also mechanical
vibrations should be suppressed by a factor of ten in future
experiments with additional passive damping systems. The deviation
at medium velocities is ascribed to molecules which do not follow
a perfect free fall trajectory, because of scattering at edges
along the beam path. In independent velocity measurements we have
already observed before the effect of scattering, which deflects
molecules of a given speed into the trajectory of another
free-fall velocity class. The red continuous line shows the
results of a model, which allows about 20\% of the molecules at
the most probable velocity to be spread out over the whole
detector area. The resulting curve then fits indeed all
experimental points, except those at low velocities, as discussed
above.

Our accumulation and imaging method requires a good mechanical
stability of the whole setup. From the good reproducibility of the
expected and observed fringe period we derive an upper limit for
the slow grating drift of 50\,nm over four hours. A drift of
10\,nm over this period is realistic in a second generation
experiment.

A clear advantage of our new detection scheme is that all velocity
classes are simultaneously recorded and encoded in the vertical
position on the screen. This ensures utmost mechanical stability
between the interferograms belonging to different velocities. The
simultaneous recording can therefore be used to measure a possible
phase shift between these interference fringes.

Ideally, we should not expect any velocity dependent phase shift
in a symmetric Talbot Lau interferometer, where the distance
between the first and second grating equals the distance between
the second and the third one~\cite{Brezger2003a}. But an
evaluation of Fig.~\ref{Fig2:Stripes} yields a phase variation in
the vertical direction of about 0.4$\pi /$mm in our experiment.
This effect can be traced back to a small angular misalignment
between the gratings around the molecular beam axis. Our
observation is for example consistent with a tilt as small as
200\,$\mu$rad between the second and the third grating.

The present detection scheme is therefore a very sensitive method
for identifying the presence of such tilts, which will be
important for interferometry with very massive molecules.
Generally, the alignment requirements increase critically with
increasing mass of the interfering particles~\cite{Stibor2005a}.

In contrast to the present setup, other grating configurations may
show additional non-classical effects, for instance the fractional
Talbot effect~\cite{Berry2001}. In particular we do expect a phase
jump of $\pi$ between fringes of certain velocity
classes~\cite{Brezger2003a} in an asymmetric Talbot Lau
configuration. This is a non-classical feature which can still be
observed in a regime where the classical Moir\'{e} effect and
quantum interference are expected to yield comparable fringe
visibilities. And the present experiment indicates that such
features should be stably recorded using our new detection method,
even in the presence of overall drifts of the interferometer.

\section{Conclusion}
Mechanically magnified fluorescence imaging offers several
advantages for future experiments aiming at recording
interferograms of nanometer-sized objects. The demonstrated method
scales favorably with the complexity of the observed particles:
organic molecules can be tagged with several dye molecules or
semiconductor nanocrystals~\cite{Bruchez1998a} and large proteins,
such as GFP~\cite{GFPbook1998a}, or again
nanocrystals~\cite{Alivisatos1996a} will even exhibit a much
higher fluorescence quantum yield and a significantly smaller
bleaching rate than the molecules in our current experiments. At
present, the smallest commercially available fluorescent
nanocrystals have a mass around
3000\,amu~\cite{Evident}\footnote{Evidenttech private
communication: The masses of the quantum dots were recently
measured by {\em Evidenttech} and turned out to be substantially
smaller than found in earlier publications.} in the core and
roughly the same mass in the ligand shell. The high efficiency of
our optical detection method will also allow to study the
relevance of different electric and magnetic dipole moments in
interference with molecules of rather similar masses, such as for
example various porphyrin derivatives. Some of them have too low
vapor pressures for experiments with ionization detectors, but
will still be detectable in fluorescence. Mechanically magnified
fluorescence imaging is therefore expected to be a scalable method
for exploring the wave-particle duality of a large class of
nanosized materials. It is an enabling technique for a range of
dephasing and decoherence studies, which will also be useful in
molecule metrology.

\section*{Acknowledgments}
This work has been supported by by the Austrian Science Fund
(FWF), within the projects START Y177 and F1505 and by the
European Commission under contract No.\,HPRN-CT-2002-00309
(QUACS). We acknowledge fruitful discussions with Klaus
Hornberger, Lucia Hackerm\"uller and Sarayut Deachapunya.

\end{document}